\documentclass[aps,prd,twocolumn,groupedaddress, nofootinbib, longbibliography,superscriptaddress,amsmath,amssymb,amsfonts, floatfix]{revtex4-2}
\usepackage{graphicx} 
\usepackage[dvipsnames]{xcolor}
\usepackage{hyperref}

\definecolor{mydarkblue}{RGB}{46, 48, 146}

\hypersetup{colorlinks=true,allcolors=mydarkblue}
\usepackage{multirow}
\usepackage{adjustbox}
\usepackage[utf8]{inputenc} 
\usepackage[T1]{fontenc}    
\usepackage{microtype}      
\usepackage{tensor}
\usepackage{orcidlink}
\usepackage{float}
\usepackage[all]{hypcap}
\usepackage{subfigure}
\usepackage[capitalise]{cleveref} 
\usepackage{url}
\usepackage{enumitem}
\usepackage{orcidlink}

\begin{document}

\title{Faster Bayesian inference with neural network bundles and new results for \texorpdfstring{$f(R)$}{f(R)} models}

\author{\mbox{Augusto T. Chantada}\orcidlink{0000-0002-4480-9595}}
\email{augustochantada01@gmail.com}
\affiliation{ Universidad de Buenos Aires, Facultad de Ciencias Exactas y Naturales, Departamento de Física, Avenida Intendente Cantilo S/N 1428 Ciudad Autónoma de Buenos Aires, Argentina}
\author{\mbox{Susana J. Landau}\orcidlink{0000-0003-2645-9197}}
\affiliation{CONICET - Universidad de Buenos Aires, Instituto de Física de Buenos Aires (IFIBA), Avenida Intendente Cantilo S/N 1428 Ciudad Autónoma de Buenos Aires, Argentina}
\author{\mbox{Pavlos Protopapas}\orcidlink{0000-0002-8178-8463}}
\affiliation{John A. Paulson School of Engineering and Applied Sciences, Harvard University, Cambridge, Massachusetts 02138, USA}
\author{\mbox{Claudia G. Scóccola}\orcidlink{0000-0002-3565-4771}}
\affiliation{Consejo Nacional de Investigaciones Científicas y Técnicas (CONICET), Godoy Cruz 2290, 1425, Ciudad Autónoma de Buenos Aires, Argentina}
\affiliation{Facultad de Ciencias Astronómicas y Geofísicas,
Universidad Nacional de La Plata, Observatorio Astronómico, Paseo del Bosque,
B1900FWA La Plata, Argentina}
\author{\mbox{Cecilia Garraffo}\orcidlink{0000-0002-8791-6286}}
\affiliation{Center for Astrophysics | Harvard \& Smithsonian, 60 Garden Street, Cambridge, Massachusetts 02138, USA}
\date{\today}

\begin{abstract}

In the last few years, there has been significant progress in the  development of machine learning methods tailored to astrophysics and cosmology. We have recently applied one of these, namely, the neural network bundle method, to the cosmological scenario. 
Moreover, we showed that in some cases the computational times of the Bayesian inference process can be reduced. In this paper, we present an improvement to the neural network bundle method that results in a significant reduction of the computational times of the statistical analysis. 
 The novel aspect consists of the use of the neural network bundle method to calculate the luminosity distance of type Ia supernovae, which is usually computed through an integral with numerical methods. In this work, we have applied this improvement to the Hu-Sawicki and Starobinsky $f(R)$ models.
We also performed a statistical analysis with data from type Ia supernovae of the Pantheon+ compilation and cosmic chronometers. 
Another original aspect of this work is the different treatment we provide for the absolute magnitude of type Ia supernovae during the inference process,  which results in different estimates of the distortion parameter than the ones obtained in the literature. We show that the statistical analyses carried out with our new method require lower computational times than the ones performed with both the numerical and the neural network method from our previous work. This reduction in time is more significant in the case of a difficult computational problem such as the ones  addressed in this work.

\end{abstract}

\maketitle

\section{Introduction}
One of the central challenges in theoretical cosmology lies in identifying the physical mechanism responsible for the current accelerated expansion of the Universe. According to the standard cosmological model $\Lambda$CDM, this acceleration can be explained by including a cosmological constant in Einstein equations.  Although this model successfully explains a broad set of existing observational data \cite{Planckcosmo2018}, it falls short in accounting for the 
value of the cosmological constant inferred from observations. Furthermore, there are some inconsistencies within the $\Lambda$CDM framework, such as the so-called Hubble tension, a disagreement between the value of the Hubble constant inferred from data of the cosmic microwave background (CMB) assuming the standard cosmological model \cite{Planckcosmo2018}, and
the one obtained from type Ia supernovae and Cepheid data \cite{Riess2022}, which are model independent. 

Motivated by this observational discrepancy, researchers explore alternative cosmological models to account for the current accelerated expansion and resolve the Hubble tension.
These alternative frameworks generally fall into two families: (i) dark energy and (ii) modified gravity. In dark energy models, a new component is added to the energy-momentum tensor, either in the form of a fluid with a time-dependent equation of state or a scalar field that is minimally coupled to gravity with a specific potential. On the contrary, modified gravity theories propose alternatives to general relativity to describe the gravitational interaction. One such example are $f(R)$ theories, which introduce modifications to the Einstein-Hilbert action in the form of a
function of the Ricci scalar. Typically, the differential systems that describe the cosmological dynamics in these modified gravity models are more complex than those of the standard $\Lambda$CDM model. Consequently, the computational times to perform parameter inference in these alternative models are significantly larger than those needed for the $\Lambda$CDM model.

Recently, neural networks (NNs) have gained traction as powerful tools for handling both data and models. In cosmology these have been employed to extract additional information from existing data in order to constrain cosmological models \cite{NN_plus_cosmo_data_exta_or_inter_1,NN_plus_cosmo_data_exta_or_inter_2,NN_plus_cosmo_data_exta_or_inter_3,NN_plus_cosmo_data_exta_or_inter_4} and to create faster Monte Carlo samplers for Bayesian inference \cite{acceletator_1, accelerator_2}. Other applications to cosmology include training  NNs as emulators of Einstein-Boltzmann solvers \cite{emulators_1, emulators_2}.
Additionally, unsupervised methods for solving differential equations have been developed \cite{NN_diff_eqs} and used in applications like physics-informed neural networks \cite{pinns}. A notable extension of these methods is the NN bundle method~\cite{bundlesolutions}, which generates a set of solutions of a given differential system. The advantage of this method is that the outcome of the NN method can be regarded as a function of the independent variable and the free parameters of the differential system. Moreover, this method does not require the use of traditional numerical solvers to obtain the solutions.

We have previously demonstrated the efficacy of this method in solving the background dynamics of the Universe for four different cosmological models \cite{our_previous_paper}. We validated the solutions obtained through the NN bundle method by comparing them with solutions from a traditional numerical solver. The main advantage of this method is that the integration of the differential system is performed only once, and thereafter the solutions can be used indefinitely for parameter inference. This reduces the computational time for statistical analysis using the Markov chain Monte Carlo (MCMC) algorithm, particularly for cosmological models that are computationally intensive to integrate.

In this work, we take a step further by demonstrating that using the NN method to compute the luminosity distance significantly reduces the time required for the inference process. 
To illustrate this, we consider the $f(R)$ Hu-Sawicki model, which we tackled in our previous work, and additionally we analyze the Starobinsky model. The latter is computationally more intensive to integrate than the former and therefore represents a more challenging test for the NN method. We test the predictions of the background dynamics of these models with recent data from cosmic chronometers \cite{CC1,CC2,CC3,CC4,CC5,CC6,CC7,CC8} and type Ia supernovae (SNIa) \cite{PantheonPlus_data}. The originality of our statistical analysis lies in the choice of the prior of SNIa absolute magnitude, which is different than the ones adopted in the literature for the same models and datasets \cite{f_R_with_custom_z_0}. This results in different values of the inferred distortion parameter $b$ of the $f(R)$ model.

The structure of this article is as follows. In Sec.~\ref{model}, we briefly describe the basics of both the Hu-Sawicki and  Starobinsky $f(R)$ models. In Sec.~\ref{method}, we describe the details of the NN bundle method when applied to 
both $f(R)$ models. We also describe the improvements to the inference process that involve the NN bundle method and leads to the optimization of computational times. 
Afterward,
we show a comparison between the solutions obtained with the NN method for the $f(R)$ models and the ones obtained with a traditional numerical solver. Additionally, we describe the appropriate treatment of the prior assumed for the absolute magnitude of SNIa.
In Sec.~\ref{Results}, we show the results of the statistical analysis where we compare the predictions of our model with data from cosmic chronometers and type Ia supernovae. We also discuss our results as compared with similar analyses \cite{f_R_with_custom_z_0, Farrugia2021}.  Most importantly, we show the significant time reduction of computational times achieved in this
study compared to both 
our previous work and the numerical method. In Sec.~\ref{conclusions}, we present our conclusions.

\section{Theoretical models} 
\label{model}

In this paper, we focus on a particular class of modified gravity theories, namely, $f(R)$ theories, whose action can be written as 
\begin{equation}
\label{S_fR}
    S = \dfrac{1}{2 \kappa}\int d^4x\sqrt{-g}f(R) + S_m.
\end{equation}
Here $\kappa =  8 \pi G$ (we adopt $c = 1$ throughout this paper), $R$ is the Ricci scalar, $g$ is the metric determinant, $f(R)$ is an arbitrary function of $R$, and $S_m$ refers to the matter action. 
From Eq.~\eqref{S_fR}, the modified Einstein equations can be derived. 

The assumption of an isotropic and homogeneous universe, which is the usual cosmological scenario, naturally leads us to use the Friedmann-Lemaître-Robertson-Walker metric.
By applying the modified Einstein equations to the cosmological setting, we obtain the following modified Friedmann equations, which describe the dynamical evolution of the background: 

\begin{eqnarray}
H^{2} &=& \dfrac{1}{3f_{R}}\left[\kappa \rho_{m}+\dfrac{Rf_{R}-f}{2}-3H\dot{R}f_{RR}\right],\nonumber\\ 
2\dot{H}+3H^{2} &=& - \dfrac{1}{f_{R}}\left[-\dfrac{Rf_{R}-f}{2}+f_{RRR}\dot{R}^{2} \right.\nonumber\\
&&+ \left. \left(\ddot{R}+2H\dot{R}\right)f_{RR}\right],
\label{friedmod}
\end{eqnarray} 

where we use the notation 
$f = f(R)$, $f_R=df/dR$, $f_{RR}=d^2f/dR^2$, etc., 
and the dot refers to the derivative with respect to cosmic time. Also, $\rho_m$ is the total matter energy density and $H$ refers to the Hubble parameter.

In this paper, we analyze the Hu-Sawicki and Starobinsky models with $n=1$. For the Hu-Sawicki model \cite{Hu-Sawicki},

\begin{equation}
    f_{\rm HS}\left(R\right)=R-2\Lambda\left[1-\dfrac{1}{1 + \dfrac{R}{\Lambda b}}\right],\label{f_R_HS_n=1}
    \end{equation}

while for the Starobinsky model \cite{Starobinsky},

\begin{equation}
f_{\rm S}\left(R\right)=R-2\Lambda\left[1-\dfrac{1}{1+\left(\dfrac{R}{\Lambda b}\right)^2}\right].\label{f_R_Staro}
\end{equation}
Here $\Lambda$ is the cosmological constant and $b$ is the distortion parameter that measures the model's departure from $\Lambda$CDM. To solve the Friedmann equations, we assume the following change of variables \cite{f_R_var, our_previous_paper}:
\begin{align}
\label{var_f_R}
x= \dfrac{\dot{R}f_{RR}}{Hf_{R}}, ~&~~~ v= \dfrac{R}{6H^{2}}, ~~~~ y= \dfrac{f}{6H^{2}f_{R}},\nonumber\\
\\
\Omega= & \dfrac{\kappa\rho_{m}}{3H^{2}f_{R}},~~~~ r=\dfrac{R}{\Lambda}. \nonumber
\end{align}

In this way, the Friedmann equations can be written as follows:

\begin{equation}
\left\{ \begin{aligned}\dfrac{dx}{dz} & =\dfrac{1}{1+z}\left(-\Omega-2v+x+4y+xv+x^{2}\right)\\
\dfrac{1}{\Gamma}\dfrac{dy}{dz} & =\dfrac{-1}{1+z}\left(vx-\dfrac{xy+4y-2yv}{\Gamma}\right)\\
\dfrac{1}{\Gamma}\dfrac{dv}{dz} & =\dfrac{-v}{1+z}\left(x+\dfrac{4-2v}{\Gamma}\right)\\
\dfrac{d\Omega}{dz} & =\dfrac{\Omega}{1+z}\left(-1+2v+x\right)\\
\dfrac{1}{\Gamma}\dfrac{dr}{dz} & =\dfrac{-r x}{1+z},
\end{aligned}
\right.\label{diff_f_R_2}
\end{equation}
where  the factor $\Gamma = \frac{f_R}{R f_{RR}}$. If we assume the Hu-Sawicki model, $\Gamma$ can be expressed as follows:

\begin{equation}
\label{Gamma_HS}
\Gamma_{\rm HS}\left(r\right)=\dfrac{\left(r+b\right)\left[\left(r+b\right)^{2}-2b\right]}{4br},
\end{equation}
while for the Starobinsky model
\begin{equation}
\label{Gamma_ST}
\Gamma_{\rm S}\left(r\right) =\dfrac{\left(r^{2}+b^2\right)\left[\left(r^{2}+b^2\right)^{2}-4 b^2 r\right]}{4 r b^2(3r^{2}-b^2)}.
\end{equation}
To facilitate the implementation of the method, an additional variable change ($r^\prime=\ln r$) is done in Eqs.~\eqref{diff_f_R_2} before the training of the NN.

The initial conditions of the system were chosen so that at high redshift the model behaves as the $\Lambda$CDM one; in this work for obtaining the solutions with the NN method we use $z_0=10$, where $z_0$ is the redshift at which the initial conditions of Eqs.~\eqref{diff_f_R_2} are set (see Sec.~\ref{subsec:acc_sol} for a discussion of the initial conditions).  For further details on the change of variables and initial conditions, we refer the reader to our previous work \cite{our_previous_paper}.

Note that the second, third, and fifth equations in the differential system in Eqs.~\eqref{diff_f_R_2} are expressed nontraditionally, that is, the factor $\Gamma$ appears in the denominator, on the left-hand side of the equation. This way of writing the differential system was used only for the Starobinsky model and the reason for this will be made clearer in Sec.~\ref{method}.

To compare the theoretical model with the observations, the Hubble parameter is needed. Therefore, after solving the system of equations described in \cref{diff_f_R_2}, one can obtain the solution for $H$ using the following relation:
\begin{equation}
    H=H^\Lambda_{0}\sqrt{\dfrac{r}{2v}\left(1-\Omega^\Lambda_{m,0}\right)}.
    \label{H_fR}
\end{equation}
Here $\Omega_{m,0}^{\Lambda}$ is the matter density parameter in the $\Lambda \mathrm{CDM}$ model which should be distinguished from $\Omega_{m,0}$, the matter density parameter defined in the $f(R)$ model. Similarly, $H^\Lambda_{0}$ denotes the Hubble constant in the $\Lambda$CDM model, while $H_0$ is the same quantity in the $f(R)$ model. 
Since the matter energy density is an observable quantity, the following relation holds:
\begin{equation}
\Omega_{m,0} H_0^2 = \Omega_{m,0}^{\Lambda} \left(H_0^{\Lambda}\right)^2.
\label{LCDMtoAlt}
\end{equation}

\section{Method}\label{sec:method}
\label{method}
\subsection{The new NN bundle method}
\label{bundle_method}
In our previous work \cite{our_previous_paper}, we extensively discussed the application of the NN bundle method in the cosmological context.  Below, we recall the main aspects of the method. First, we describe the method for the case in which neither the differential system nor the initial condition depends on any parameter. The NN  method can be formulated as an optimization problem. The goal is to tune the networks' internal parameters to minimize a specific loss function. We denote the vector of the outputs of the NNs as $\boldsymbol{u}_{\mathcal{N}}\left(t\right)$, where $t$ is the independent variable. The loss function to be minimized is then\footnote{Actually, the  loss function in Eq. \eqref{loss_res_reparam} is evaluated at $N_{\rm batch}$ random points of the domain in each iteration of the optimization process so that the total loss at each iteration is  $J=\frac{1}{N_{\rm batch}}\sum_i^{N_{\rm batch}} L\left(\tilde{\boldsymbol{u}}(t_i),t_i\right)$.}
\begin{equation}
    L\left(\tilde{\boldsymbol{u}},t\right)=\sum_{i}^{M}\mathcal{R}_{i}\left(\tilde{\boldsymbol{u}},t\right)^{2},
    \label{loss_res_reparam}
\end{equation}
where $\mathcal{R}_{i}$ are the residuals of the $M$ differential equations, i.e. the left-hand side of the $i$th differential equation minus its right-hand side,\footnote{It follows that, if $\tilde{\boldsymbol{u}}$ is the exact solution, it satisfies the differential system and therefore the loss function is exactly zero. Since the outcomes of the NN method are not exact solutions, the loss measures how far the solutions are from satisfying the differential system. For details, see Ref.~\cite{our_previous_paper}.} 
 and $\tilde{\boldsymbol{u}}\left(t\right)$ is a reparametrization of the outputs $\boldsymbol{u}_{\mathcal{N}}\left(t\right)$ so as to enforce the initial condition $\boldsymbol{u}_0$,
\begin{equation}\tilde{\boldsymbol{u}}\left(t\right)=\boldsymbol{u}_0 +\left(1-e^{-\left(t-t_0\right)}\right)\boldsymbol{u}_{\mathcal{N}}\left(t\right).
\end{equation}
Now, we move to the NN bundle method, which is applied when the differential system and/or the initial conditions depend on free parameters. We will call  $\boldsymbol{\theta}$ the vector that includes all those parameters. In this case, the loss function to be optimized reads\footnote{Actually, the  loss function in Eq. \eqref{loss_res_reparam_bundle} is evaluated at $N_{\rm batch}$ random points of the domain in each iteration of the optimization process so that the total loss at each iteration is  $J=\frac{1}{N_{\rm batch}}\sum_i^{N_{\rm batch}} L\left(\tilde{\boldsymbol{u}}, t_i,\boldsymbol{\theta}_i\right)$.}

\begin{equation}
    L\left(\tilde{\boldsymbol{u}},t,\boldsymbol{\theta}\right)=\sum_{i}^{M}\mathcal{R}_{i}\left(\tilde{\boldsymbol{u}},t,\boldsymbol{\theta}\right)^{2},
    \label{loss_res_reparam_bundle}
\end{equation}
and $\tilde{\boldsymbol{u}}\left(t, \boldsymbol{\theta}\right)$ is a reparametrization of the outputs $\boldsymbol{u}_{\mathcal{N}}\left(t, \boldsymbol{\theta}\right)$ so as to enforce the initial condition $\boldsymbol{u}_0(\boldsymbol{\theta})$,

\begin{equation}\tilde{\boldsymbol{u}}\left(t,\boldsymbol{\theta}\right)=\boldsymbol{u}_0 (\boldsymbol{\theta})+\left(1-e^{-\left(t-t_0\right)}\right)\boldsymbol{u}_{\mathcal{N}}\left(t,\boldsymbol{\theta}\right).
\end{equation}
In this way, when the optimization process finishes, the trained neural network represents the solution of the differential system for any value of the free parameters $\boldsymbol{\theta}$ within the training range. This is different from numerical methods, where the system needs to be integrated for every choice of the free parameters.

With regards to the unsupervised nature of the training, the key to understanding it lies in the residuals, which are calculated from the differential equations of the system. In this way, each residual depends on the independent variable, the bundle parameters, the evaluated reparametrized networks, and nothing else. At each iteration, the residuals are evaluated at random points of the domain (independent variable and bundle parameters). This calculation involves evaluating the reparametrized networks and their derivatives\footnote{These are not to be confused with derivatives involving the stochastic gradient descent algorithm used to optimize the networks. The derivatives mentioned here are the ones necessary to compute the residuals, i.e., the derivatives that appear in the differential equations.} (with respect to the independent variable) at these points. Then, the value of the loss is calculated, and the internal parameters of the network are tuned in order to lower it.

\begin{figure}
    \centering
    \includegraphics[scale=0.83]{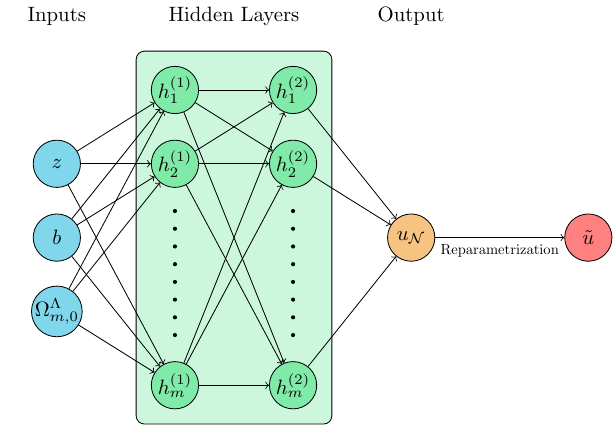}
    \caption{Graph that represents a neural network like the ones employed in this work. It has three inputs that correspond to the independent variable and the two parameters of the bundle, two hidden layers with the same amount of units each, and a single output that is then reparametrized to satisfy the initial condition.}
    \label{fig:NN_f_R_diagram}
\end{figure}

To implement the method to solve the background equations of the $f(R)$ models, we use the loss described in Eq. \eqref{loss_res_reparam_bundle}.\footnote{We also add an extra term to the loss in Eq.~\eqref{loss_res_reparam_bundle} to explicitly enforce relationships between the dependent variables that must be satisfied (see Appendix B 3 of our previous work \cite{our_previous_paper} for more information).}
When the solution of a given differential system is known for a fixed value of one or more of the free parameters, it is useful to consider the perturbative reparametrization that was proposed and applied in our previous work \cite{our_previous_paper} to the quintessence and $f(R)$ Hu-Sawicki models. When applied to the $f(R)$ models that are concerned here, this 
takes the form

\begin{equation}
\begin{aligned}
\tilde{\boldsymbol{u}}\left(z, b, \Omega_{m,0}^{\Lambda}\right)&=\hat{\boldsymbol{u}}\left(z,\Omega_{m,0}^{\Lambda}\right) +\left(1-e^{-\left(z-z_0\right)}\right)\times\\ & \left(1-e^{-b}\right)\boldsymbol{u}_{\mathcal{N}}\left(z, b, \Omega_{m,0}^{\Lambda} \right).
\end{aligned}
\label{perturbative_reparam_bundle_f_R}
\end{equation}
Here $\hat{\boldsymbol{u}}\left(z,\Omega_{m,0}^{\Lambda}\right)$ denotes the $\Lambda$CDM  solution of Eqs.~\eqref{diff_f_R_2} (i.e. for $b\xrightarrow{}0$) which is already known, and the job of the second term is to correct it in order to make the solution valid for the whole bundle. We recall that the bundle parameters are $b$ and $\Omega_{m,0}^\Lambda$.\footnote{The parameter $b$ enters in Eqs. \eqref{diff_f_R_2} through  $\Gamma$ [Eqs. \eqref{Gamma_HS} and \eqref{Gamma_ST}], while the initial conditions depend on $\Omega_{m,0}^\Lambda$ (see our previous work \cite{our_previous_paper}).}In Fig.~\ref{fig:NN_f_R_diagram} we show a graph that represents the neural network described before.

This reparametrization helps to circumvent the singularity of the differential system at $b=0$ [see Eqs.~\eqref{Gamma_HS} and \eqref{Gamma_ST}]. In contrast, for numerical methods, an approximate analytical function is usually used when $b$ approaches $0$, since the singularity causes an increase in the computational times of the statistical analysis \cite{Basilakos}.

Regarding the implementation of the method to the Starobinsky model, it is important to highlight our approach in formulating Eqs.~\eqref{diff_f_R_2}. Traditional numerical solvers typically require differential equations to be arranged with the derivative of the independent variable isolated on one side and the rest on the other side. However, the NN method is not bound by this constraint; the residual can be structured in any way that preserves the same equality that the equation conveys.
Thus, we chose to define the residuals based on the formulation presented in Eqs.~\eqref{diff_f_R_2}, inspired by the fact that the singularity in $\Gamma$ makes the balance between the equations uneven. 
Consequently, we use the residuals that correspond to the second, third, and fifth equations of Eqs.~\eqref{diff_f_R_2}. Therefore, these residuals no longer exhibited a divergent behavior close to $b=0$.
This shows a useful flexibility that the NN method has that numerical methods do not. This last technique was needed in addition to the perturbative reparametrization to make the solutions accurate at low values of $b$ for the Starobinsky model. 
In the case of the Hu-Sawicki model, the differential equations and, in consequence, the residuals and the loss are written in the usual way (see our previous paper for details \cite{our_previous_paper}).

In this work, we present an improvement that significantly reduces the time it takes to perform an MCMC analysis. This improvement is related to the computation of the luminosity distance of type Ia supernovae,
\begin{equation}
    d_L\left(z\right)=\left(1+z\right)\int^z_0\dfrac{dz^\prime}{H\left(z^\prime\right)}.
    \label{dl_theo}
\end{equation}

Building upon the previous equation, it becomes evident that we must carry out an integration involving an integrand $ \frac{1}{H}$, derived from the output of the NN method. In the context of numerical integration methods, the integrand (which in this case is a function of the NN) must be evaluated at $N_{\text{int}}$ different values of $z$. The computational time for this process is directly proportional to $N_{\text{int}}$. Furthermore, it is important to note that the accuracy of the numerical integration method improves with higher values of $N_{\text{int}}$. Therefore, this poses a computational bottleneck, particularly when this integral must be recalculated at each parameter sample in the MCMC process.

To mitigate this computational challenge, one can consider NNs once again to evaluate the integral. Using the fundamental theorem of calculus, we can transform the integral equation into a differential equation. This way, we define

\begin{equation}
\begin{aligned}
\label{eq:integral_to_diffeq}
        \dfrac{dI}{dz}=\dfrac{1}{E\left(z\right)}, &&
    \left.I\left(z\right)\right|_{z=0}=0,
\end{aligned}
\end{equation}
where $E\left(z\right)=H\left(z\right)/H_0^\Lambda$ and $I\left(z\right)=d_L\left(z\right) H_0^\Lambda/\left(1+z\right)$. 
Next we describe the specific steps of our new NN method. First, a set of NNs are trained to obtain $E\left(z, b,\Omega^\Lambda_{m,0}\right)$ from Eqs.~\eqref{diff_f_R_2}. The trained NNs are then used in Eq.~\eqref{eq:integral_to_diffeq} to define another differential equation which is also solved with the NN bundle method. Therefore,  after training, we obtain a new NN that represents $I\left(z, b, \Omega^\Lambda_{m,0}\right)$. 
In this way, the integration that is needed to obtain
the luminosity distance is no longer computed during the MCMC process. Instead, it is performed in the NN training that yields $I\left(z, b, \Omega^\Lambda_{m,0}\right)$ (i.e., a bundle solution of the integral), resulting in a reduction of the computational times of the inference process. We also note that in this way, the likelihood is completely analytical since all of the theoretical predictions of the observable quantities 
come from NNs that work as functions of the parameters and the independent variable. 
In short, with our new approach the NN bundle method is applied twice: first to solve the differential system and second to compute the luminosity distance. 
On the contrary, in our previous work~\cite{our_previous_paper}, we only used the  NN bundle method to solve the differential system, while the luminosity distance was computed with numerical methods.

The concept of using NNs to perform an integration has been explored by other authors \cite{NN_int_analytical, integration_paper_1, integration_paper_2, 2023_integration}. Specifically, the idea of using the fundamental theorem of calculus to turn an integral equation into a differential one and after that use a NN-based method to solve it has been explored in Refs.~\cite{integration_paper_1, integration_paper_2, 2023_integration}. Nevertheless, our work is the first one to extend this concept to the case where the integrand is a solution previously obtained by the NN bundle method. Indeed, the integrand of Eq.~\eqref{eq:integral_to_diffeq} is computed from the solution of the differential system described by Eqs.~\eqref{diff_f_R_2}.

\subsection{Accuracy of solutions}\label{subsec:acc_sol}

All of the NNs in this work have three inputs (corresponding to $z$, $b$, and $\Omega^{\Lambda}_{m,0}$), two hidden layers of 32 units each, and 1 output for a total of 1217 trainable parameters per NN. 
 Each NN has a single output so that, when reparametrized as in Eq. \eqref{perturbative_reparam_bundle_f_R},
 it should approximate the associated dependent variable that solves the
corresponding differential equation. It should be noted that the specific 
architecture used in this work was chosen based on its simplicity and 
effectiveness. 
 Indeed, it has proven to be effective with a relatively low amount of trainable parameters.

The training ranges for the input parameters of the NNs are as follows: 
for the Hu-Sawicki model,  $z\in\left[0,10\right]$, $b\in\left(0,3\right]$, and $\Omega_{m,0}^\Lambda\in\left[0.05,0.4\right]$, while for the Starobinsky model, they are $z\in\left[0,10\right]$,
$b\in\left(0,4\right]$, and $\Omega_{m,0}^\Lambda\in\left[0.1,0.4\right]$.

When the training is finished, the NN internal parameters adopt the values that correspond to the lowest value of the total loss. This latter value resulted to be $L_{\rm min}=1.14\times 10^{-5}$ [for Eqs.~\eqref{diff_f_R_2}] and $L_{\rm min}=5.43\times 10^{-9}$ [for Eq.~\eqref{eq:integral_to_diffeq}] in the Hu-Sawicki model, while they turned out to be $L_{\rm min}=2.19\times 10^{-6}$ [for Eqs.~\eqref{diff_f_R_2}] and $L_{\rm min}=8.16\times 10^{-9}$ [for Eq.~\eqref{eq:integral_to_diffeq}] in the Starobinsky model.

However, the training of the NNs does not give any information about the error in the quantities that we are interested in ($H$ and $d_L$). For this reason, we evaluate the precision of the outputs given by the NNs by computing the absolute value of the relative difference between these NN-based solutions and the ones obtained from a numerical solver. In particular, we used the Runge-Kutta methods\footnote{We used RKF45 for the Hu-Sawicki model and Radau IIA family of order 5 for the Starobinsky model. The latter method is meant for stiff problems and it was necessary to use it for the convergence of the Starobinsky model.} found in SciPy's \cite{SciPy} tool to solve initial value problems and consider low tolerance for both relative and absolute errors, $10^{-7}$ and $10^{-12}$, respectively. We chose the Runge-Kutta method for being well-established and tested, as well as for its presence in a widely used library.

\begin{figure*}
    \centering
    \subfigure[]{\includegraphics[width=\columnwidth]{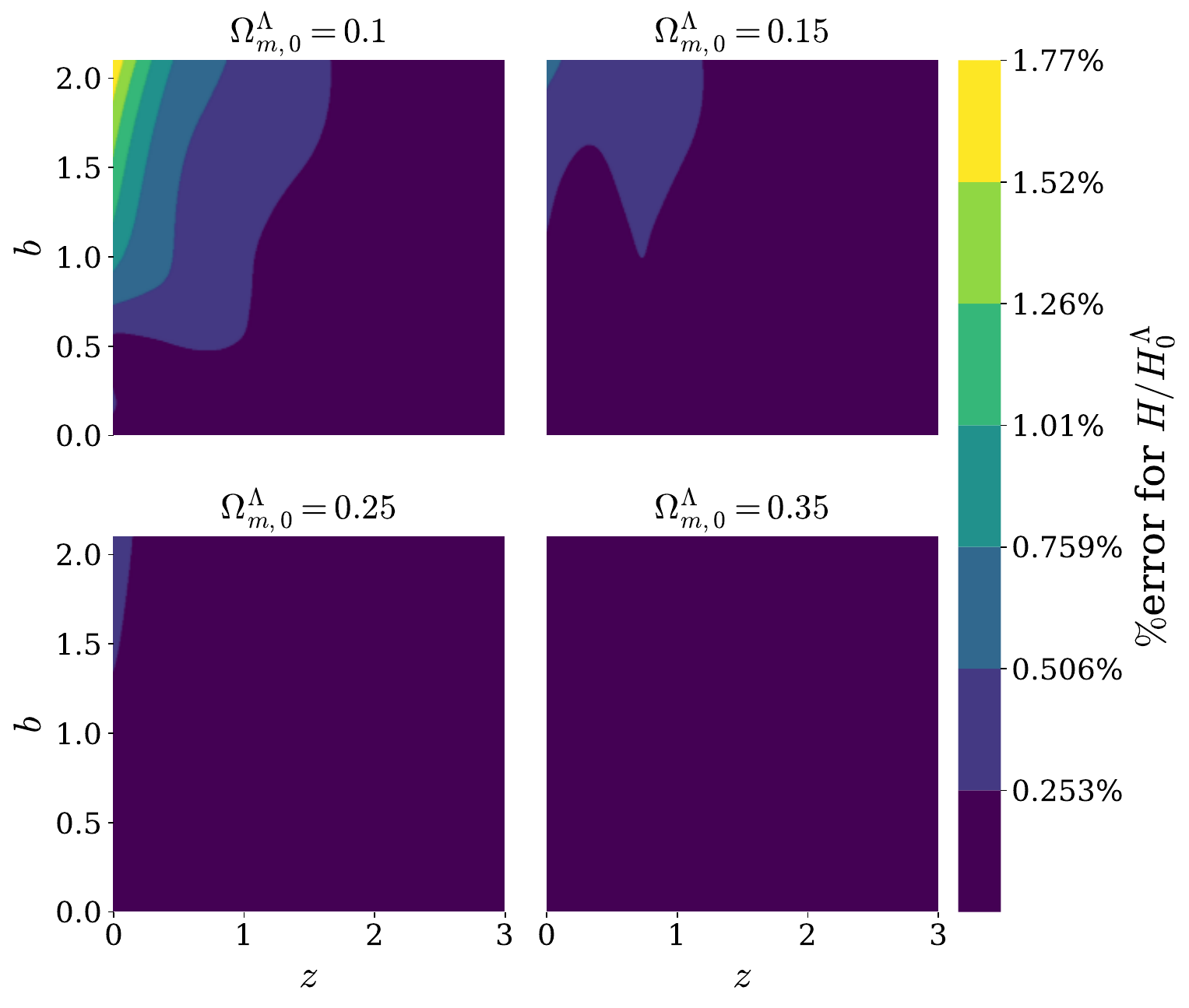}} 
    \subfigure[]{\includegraphics[width=\columnwidth]{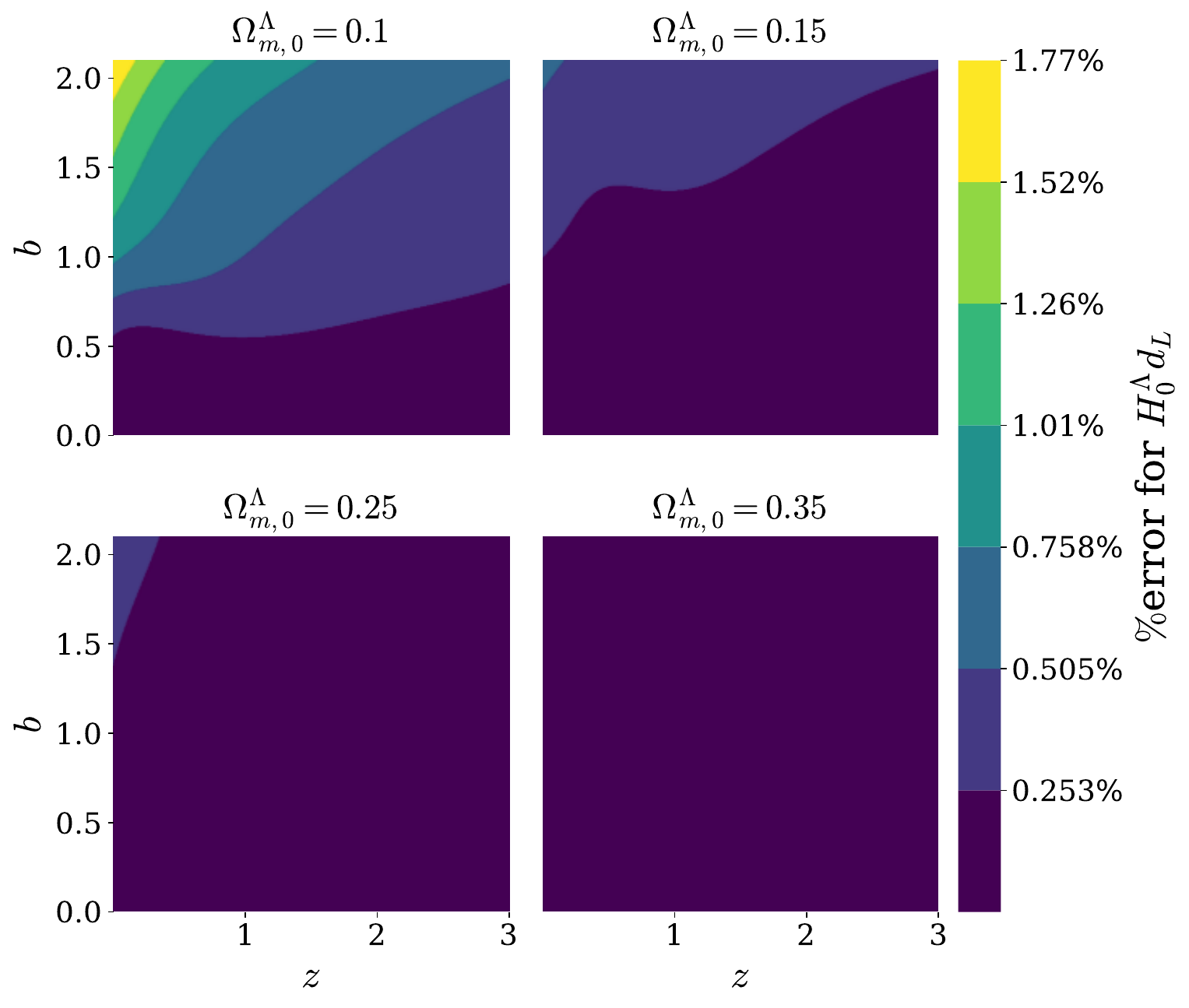}} 
    \caption{The percentage error of the NN-based bundle
solution of $H/H_0^\Lambda$ (a) and $H_0^\Lambda d_L$ (b) in the Hu-Sawicki model with $n=1$, through comparison to numerical solutions. The range of the comparison
goes through a section of the training range of the parameters
of the bundle.}
    \label{fig:heatmap_Hu-Sawicki}
\end{figure*}

\begin{figure*}
    \centering
    \subfigure[]{\includegraphics[width=\columnwidth]{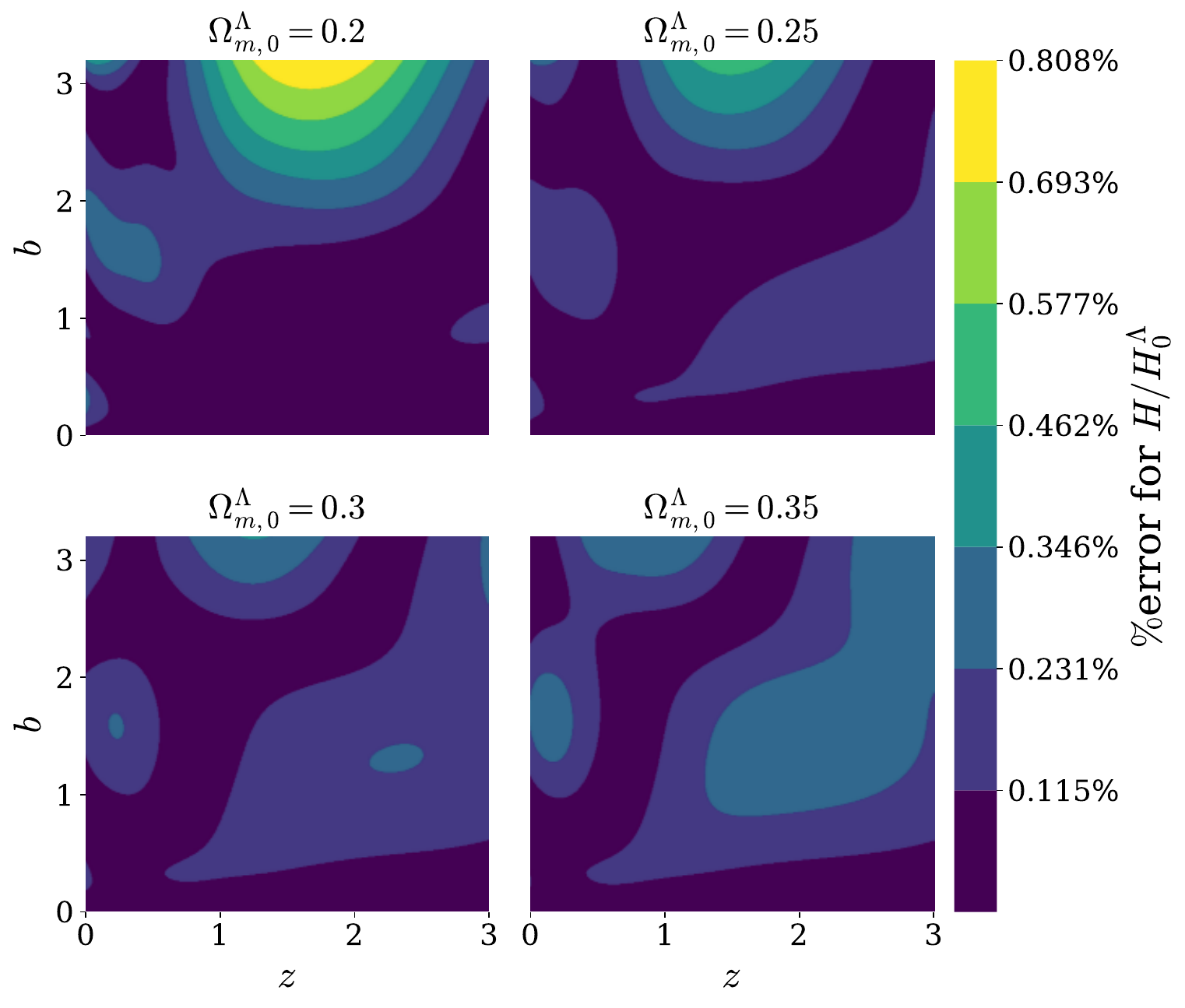}} 
    \subfigure[]{\includegraphics[width=\columnwidth]{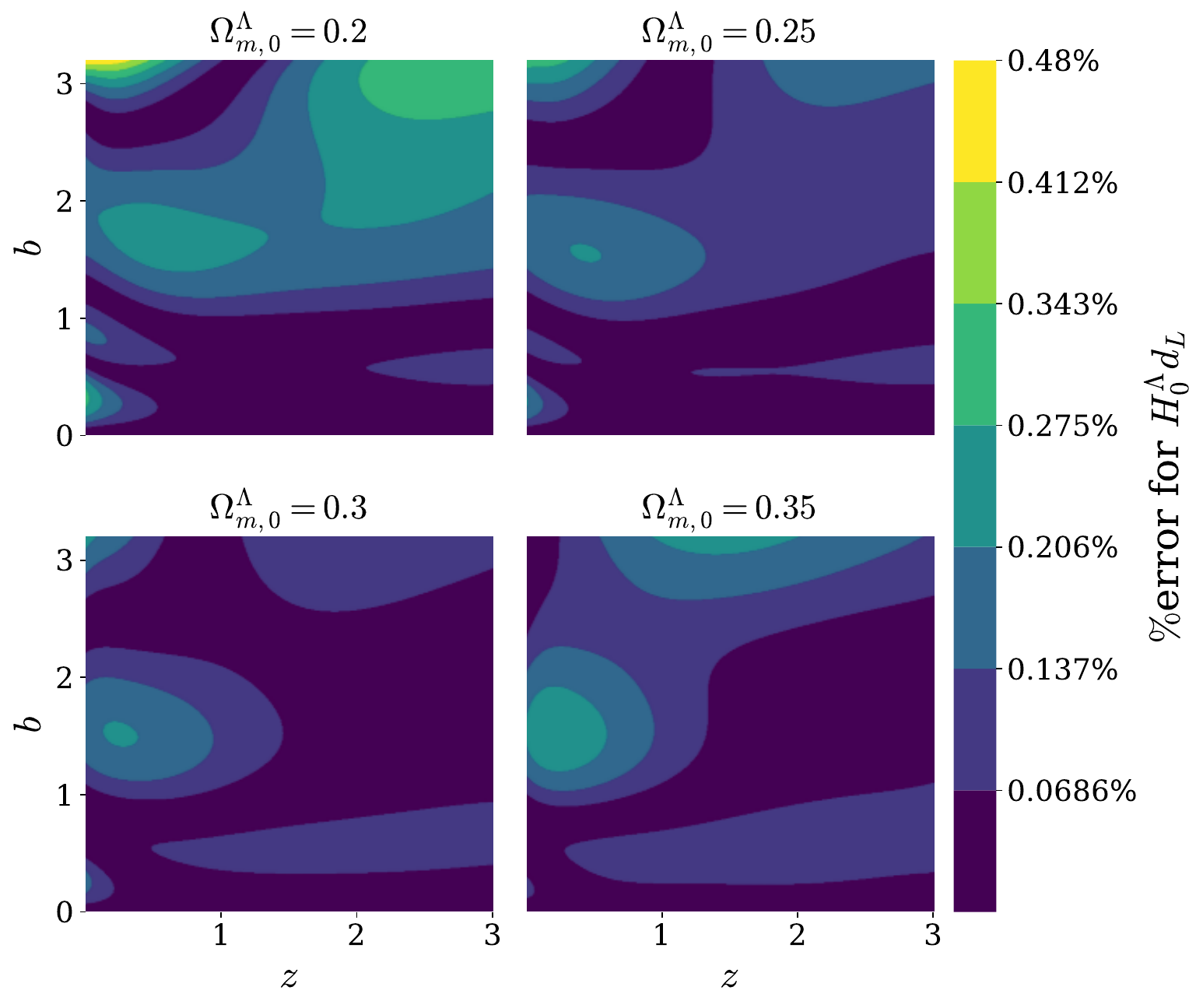}} 
    \caption{The percentage error of the NN-based bundle
solution of $H/H_0^\Lambda$ (a) and $H_0^\Lambda d_L$ (b) in the Starobinsky model with $n=1$, through comparison to numerical solutions. The range of the comparison
goes through a section of the training range of the parameters
of the bundle.}
    \label{fig:heatmap_Starobinsky}
\end{figure*}

Now, we focus on two specific quantities, $H/H_0^\Lambda$ and $H^\Lambda_0 d_L$.  The results are shown in Figs.~\ref{fig:heatmap_Hu-Sawicki} and \ref{fig:heatmap_Starobinsky}, where the range of the parameters covers the region of the parameter space that is within the $ 95 \%$ confidence level, $\left(b, \Omega^\Lambda_{m,0}\right)\in \left(0, 2.1\right]\times \left[0.1, 0.35\right]$ and $\left(b, \Omega^\Lambda_{m,0}\right)\in \left(0, 3.2\right]\times \left[0.2, 0.35\right]$ for Hu-Sawicki and Starobinsky models, respectively, as we will later show in Sec.~\ref{subsec:statistical_analysis}. 
The figures also show the dependence with redshift in the region relevant to the datasets we are considering. These plots show that the errors are, at most, $\sim 1.8 \%$ and $\sim 0.8 \%$ in the region of the parameter space evaluated, for Hu-Sawicki and Starobinsky models, respectively.

The choice of the initial condition deserves a detailed explanation. The bottom line is that we used $z_0=10$ for the NN method for both models, but in the case of the numerical method, we could only use this value for the Hu-Sawicki model, while for the Starobinsky model we needed to compute the initial condition more carefully, as we explain in what follows.

Specifically, Ref.~\cite{f_R_with_custom_z_0} highlights that using a fixed and low value of $z_0$ can lead to numerical instabilities in some areas of the parameter space. 
We verified by ourselves that there are instabilities in the numerical method in specific regions of the parameter space when using the initial condition $z_0 = 10$ in the case of the Starobinsky model.
On the other hand,
a more thorough method for computing $z_0$ was developed in Refs.~\cite{Odintsov2017,Farrugia2021}, where its value depends on both $b$ and $\Omega^{\Lambda}_{m,0}$. For the Starobinsky model with $n=1$, this is
\begin{equation}
\label{z0}
z_0=\left[\dfrac{1-\Omega_{m,0}^\Lambda - \Omega_{r,0}^\Lambda}{\Omega_{m,0}^\Lambda}\left(b\sqrt{\dfrac{1}{\epsilon} - 1} - 4 \right)\right]^{1/3}-1 
\end{equation}
with $\Omega_{r,0}^\Lambda=2.97 \times 10^ {-4} \Omega_{m,0}^\Lambda$, and $\epsilon \ll 1$.\footnote{The value of $\epsilon$ is usually determined by imposing $f\left[R\left(z=z_0\right)\right]=R-2\Lambda\left(1-\epsilon\right)$, thus it depends on the specific $f(R)$ being used \cite{Farrugia2021}. We use the values used in Ref.~\cite{f_R_with_custom_z_0}.}
This alternative method can lead to significantly higher values of $z_0$, sometimes on the order of $100$. For such high $z_0$ values, the radiation component of the Universe can no longer be neglected in the Friedmann equations and, consequently, in the initial conditions. Also, a high value of  $z_0$  results in an increase in the computing times of the solution. 
Nevertheless, using Eq.~\eqref{z0} to determine the initial condition resolved the numerical issues, and therefore we used the initial condition taken from Eq.~\eqref{z0} in all calculations performed with the numerical method for the Starobinsky model. We stress that for the Hu-Sawicki model we always use $z_0=10$. 
This discussion is relevant to understand the comparison between the results of the inference process obtained using the numerical method with those obtained using NNs (as shown later) and for the estimation of the computing times of both methods.

Going back to Fig~\ref{fig:heatmap_Starobinsky}, which displays the comparison between the NN and the numerical solutions for the Starobinsky model, we notice that the solutions obtained with the NN method using the initial condition at $z_0=10$ closely match those obtained using the numerical method, which employs Eq.~\eqref{z0}. This agreement is observed within the parameter space region encompassing the $95\%$ confidence level. Hence, one advantage of the NN method is its ability to achieve stability without requiring a more comprehensive initial condition.
A comparison between $z_0=10$ and $z_0=z_0\left(b, \Omega^\Lambda_{m,0}\right)$ with the numerical method revealed no significant differences, with variations remaining below $\sim1.74\%$ for the Hu-Sawicki model and $\sim0.06\%$ for the Starobinsky model in the same parameter space region as the one shown in Figs.~\ref{fig:heatmap_Hu-Sawicki} and \ref{fig:heatmap_Starobinsky}. However, due to stability issues during the computations of the MCMC process, it was required to use the more complex initial condition for the Starobinsky case when using the numerical method.

\subsection{Statistical analysis}\label{subsec:statistical_analysis}
\subsubsection{Datasets}
In this section we briefly describe the datasets used in the statistical analysis of this work: cosmic chronometers and SNIa. 

We use 32 measurements \cite{CC1,CC2,CC3,CC4,CC5,CC6,CC7,CC8} of $H$ that were obtained using  the cosmic chronometers technique within a range of redshift $z\in \left[0.07, 1.965\right]$. 
The likelihood function is

\begin{equation}
{\cal L}_{\bf \rm CC} \propto \exp \left\{ -\dfrac{1}{2}\sum_{i=1}^{32} \left[\dfrac{H^{\rm obs}\left(z_i\right) - H^{\rm th}\left(z_i,b, \Omega^\Lambda_{m,0}\right)}{\sigma_{H^{\rm obs}\left(z_i\right)}}\right]^2 \right\},
\label{like_CC}
\end{equation}
where $H^{\rm obs}\left(z_i\right)$ refers to the observational measurements of $H\left(z\right)$ and $\sigma_{H^{\rm obs}\left(z_i\right)}$ refers 
to the corresponding error, while $H^{\rm th}\left(z_i,b, \Omega^\Lambda_{m,0}\right)$ is the model’s prediction that is obtained by evaluating two trained NNs [one corresponding to $v\left(z_i,b, \Omega^\Lambda_{m,0}\right)$ and another to $r\left(z_i,b, \Omega^\Lambda_{m,0}\right)$\footnote{We recall that the actual dependant variable used is $r^\prime=\ln r$.} in Eq.~\eqref{H_fR}]. We emphasise that these two NNs are trained together with another three, representing the five variables in Eq.~\eqref{var_f_R}, to solve Eqs.~\eqref{diff_f_R_2}.

It is widely accepted that SNIa can be considered as standard candles due to the homogeneity of their light curves. This makes them ideal for determining distances and also to constrain cosmological parameters. 
From the  supernovae light curves, it is possible to infer the distance moduli $\mu$, which can be expressed as
\begin{equation}
\label{mu_abrev}
    \mu_{\rm obs} = \tilde{m}_b - M, 
\end{equation}
where $\tilde{m}_b$ refers to the corrected overall flux normalization (see \cref{nuissance}) and $M$ is the absolute magnitude. This last equation is a simplified version of the Tripp formula \cite{Tripp}, which includes additional corrections described  by extra parameters that will not be used in this work. In \cref{nuissance}, we describe the complete Tripp formula and the reasons that lead us to use Eq.~\eqref{mu_abrev} in this work. 
On the other hand, the distance moduli are related to the luminosity distance [Eq.~\eqref{dl_theo}] as
\begin{equation}
    \label{distance_moduli}
    \mu\left(z\right) = 25 + 5\log_{10}\left(d_L\left(z\right)\right).
\end{equation}
In this work, we consider 1590 SNIa, at redshifts $z\in\left[0.01, 2.26\right]$ from the 
Pantheon+ compilation \cite{PantheonPlus_data}.

The likelihood function can be written as
\begin{equation}
{\cal L}_{\bf \rm SNIa} \propto \exp \left\{ -\dfrac{1}{2} \Delta \boldsymbol{\mu}^T \cdot\boldsymbol{C}^{-1}\cdot \Delta \boldsymbol{\mu}\right\},
\label{like_SN}
\end{equation}
where the vector $\Delta \boldsymbol{\mu} = \boldsymbol{\mu}^{\rm obs} - \boldsymbol{\mu}^{\rm th}\left(b, \Omega^\Lambda_{m,0}\right)$ contains the difference between the observed value and the theoretical prediction 
of the distance modulus for each measurement in the compilation, while $\boldsymbol{C}$ denotes the covariance matrix. The theoretical predictions of $\mu$ are computed using a single NN corresponding to $I\left(z,b,\Omega^\Lambda_{m,0}\right)$ in combination with the fact that $d_L=I\left(1+z\right)/H_0^\Lambda$ and Eq.~\eqref{distance_moduli}. This last NN is only trained after the training associated with solving Eqs.~\eqref{diff_f_R_2} is done. Then, using the resulting trained NNs, Eq.~\eqref{eq:integral_to_diffeq} can be defined to now train the NN that represents $I\left(z,b,\Omega^\Lambda_{m,0}\right)$.

\subsubsection{ Methodology for the statistical analysis}
\label{subsubsec: methodology_stat}
The likelihood function used in the MCMC process is the product of two separate likelihood functions: the one in Eqs.~\eqref{like_CC} and the one in Eq.~\eqref{like_SN}. We recall that the free parameters of the posterior distribution are $b$, $\Omega_m^\Lambda$, $H_0^\Lambda$, and $M$. We use uniform priors for $b$, $\Omega^\Lambda_{m,0}$, and $H_0^\Lambda$; $b \in (0,3]$, $\Omega^\Lambda_{m,0} \in [0.05,0.4]$, and $H_0^\Lambda \in [50, 100]$, respectively for the Hu-Sawicki model. For the Starobinsky model we use the same prior for $H^\Lambda_0$, and the flat priors for the other two are $b \in (0,4]$ and $\Omega^\Lambda_{m,0}\in \left[0.1, 0.4\right]$.

For the treatment of the absolute magnitude of SNIa, various approaches exist in the literature. These can be summarized as (i) using a uniform prior and marginalizing $M$ \cite{2011ApJS..192....1C,f_R_with_custom_z_0}, (ii) taking $M$ as a free parameter with a uniform prior \cite{Matias_paper,our_previous_paper}, and (iii) taking $M$ as a free parameter and using a Gaussian prior with mean and standard deviation determined from independent data. We choose the (iii) method as developed in~\cite{camarena_marra_technique,camarena_marra_pantheon_plus} where the Cepheid data from the SH0ES Collaboration are used to determine the prior parameters of $M$. In particular,  we used a Gaussian distribution centered on $\mu_M=-19.243$ with $\sigma_M=0.32$. We stress that the methods described in (i) and (ii) assume equal probability of all values of $M$, which would be reasonable if no further information about this quantity is available. However, as is well known, Cepheid data from the SH0ES collaboration can be used to extract information about $M$ ~\cite{camarena_marra_technique,camarena_marra_pantheon_plus}. Therefore, our treatment is more appropriate than the one reported in Ref.~\cite{f_R_with_custom_z_0}. On the other hand, leaving $M$ as a free parameter or performing the marginalization with a uniform prior can bias the estimation of cosmological parameters.\footnote{We notice that the marginalization proposed in \cite{2011ApJS..192....1C} is performed over the variable $\mathcal{M}=M+5 \log(\frac{c/H_0}{\rm Mpc})$, which involves the Hubble constant. Therefore, this marginalization is not recommended to estimate  $H_0$ to address the Hubble tension.}
In fact, we did perform the inference process considering the 
approaches (i) and (ii) described before, and we found a significant 
change in the estimated values of $b$. On the other hand, it has been shown in Ref.~\cite{camarena_marra_pantheon_plus} that considering the Gaussian prior in this way yields very similar results to the case when the Pantheon+ and SH0ES
 data are considered. In this way, our approach allows us to include the
 information provided by the SH0ES Collaboration that is relevant for 
the Hubble tension while providing a prior on $M$ that is independent of the dataset considered for the statistical analysis.

\section{Results}
\label{Results}

In this section, we present and discuss the results of the statistical analysis using the MCMC algorithm with the NN-based solutions. We then compare the time required for the MCMC analysis using this new method with the one required by the numerical method and the method employed in our previous work~\cite{our_previous_paper}.

\subsection{Results from the statistical analysis}
\begin{figure}
    \centering
    \includegraphics[width=\columnwidth]{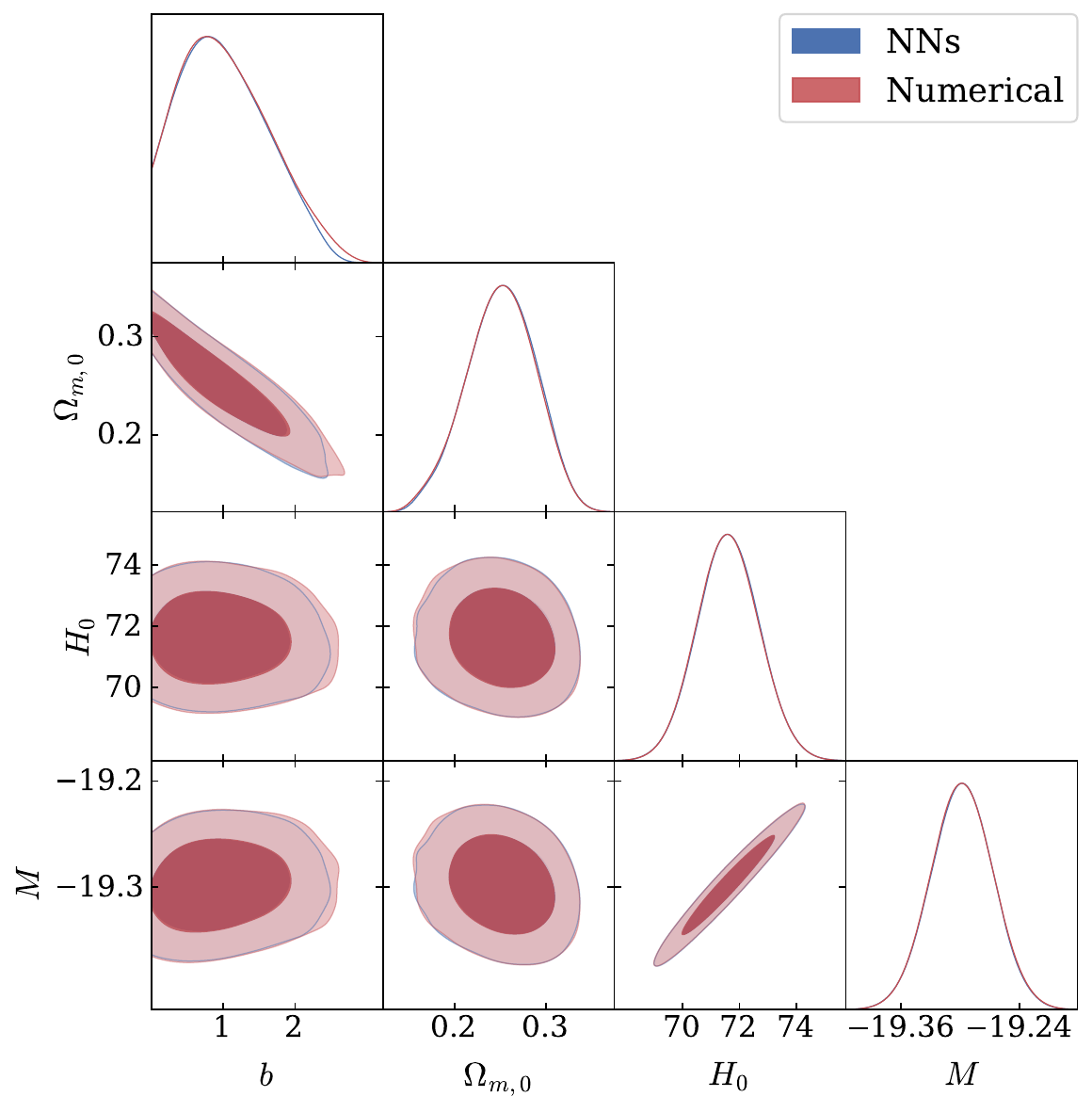}
    \caption{Results of the statistical analysis for the $f(R)$ Hu-Sawicki model using NNs (in blue) and using a numerical solver (in red). The darker and brighter regions correspond to the 68\% and 95\% confidence regions, respectively. The plots on the diagonal show the posterior probability density for each of the free parameters of the model.}
    \label{contours_Hu-Sawicki}
\end{figure}

\begin{figure}
    \centering
    \includegraphics[width=\columnwidth]{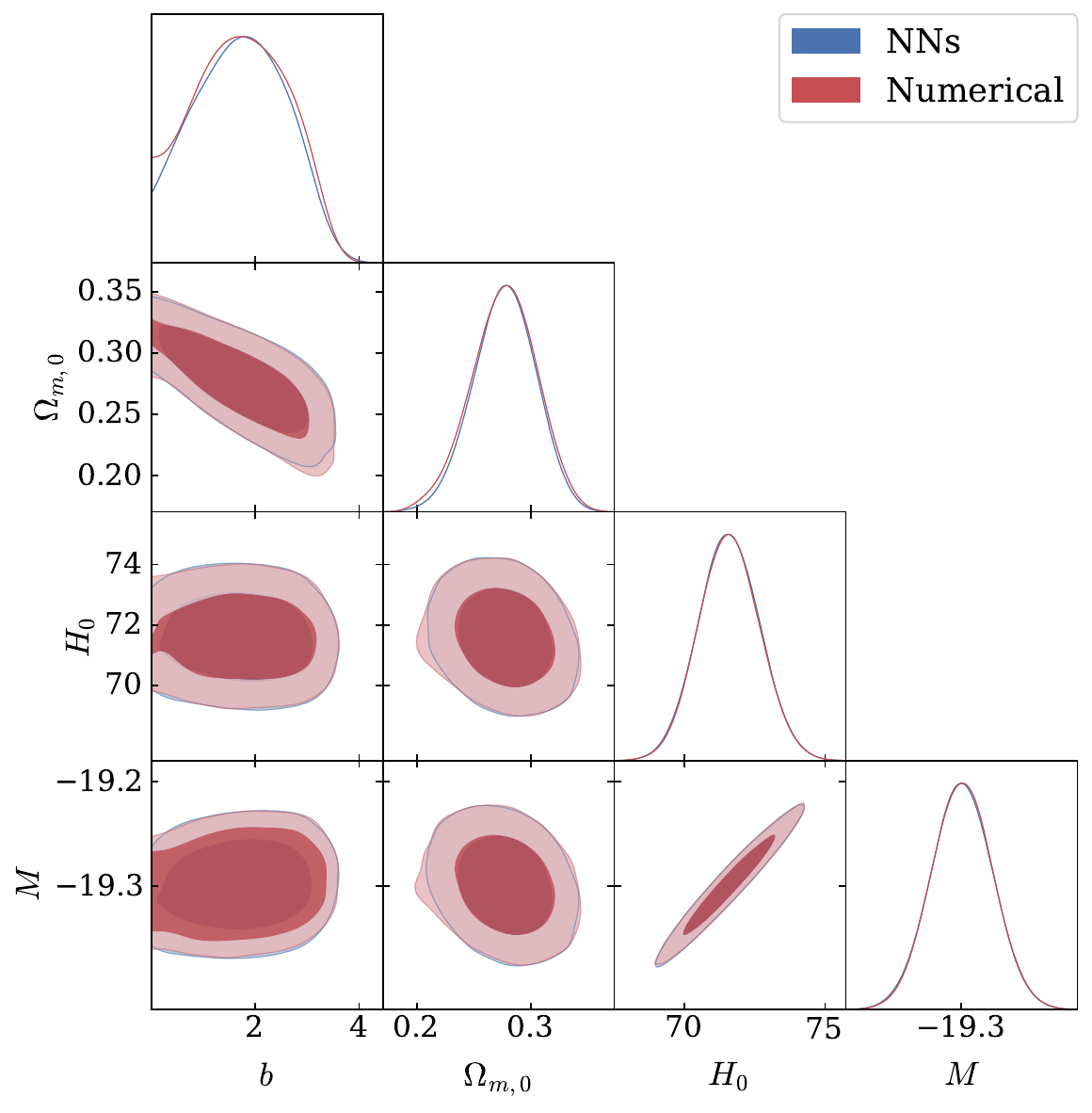}
    \caption{Results of the statistical analysis for the $f(R)$ Starobinsky model using NNs (in blue) and using a numerical solver (in red). The darker and brighter regions correspond to the 68\% and 95\% confidence regions, respectively. The plots on the diagonal show the posterior probability density for each of the free parameters of the model.}
    \label{contours_Starobinsky}
\end{figure}

In Figs.~\ref{contours_Hu-Sawicki} and ~\ref{contours_Starobinsky} , we show the 68\% and 95\% confidence level contours of the posterior probability distribution of the parameters, using a numerical method and using the NN-based solutions for both $f(R)$ models analyzed in this work. As it can be seen in the figures, the results are fully compatible with each other, showing that the errors shown in Figs.~\ref{fig:heatmap_Hu-Sawicki} and \ref{fig:heatmap_Starobinsky} do not significantly affect the results of the statistical analysis.
We also show the resulting $68 \%$ and $95 \%$ confidence intervals and mean and best fit values for the parameters of the models obtained using NNs in Tables~\ref{table:mcmc_results_HS} and~\ref{table:mcmc_results_ST} for Hu-Sawicki and Starobinsky models, respectively. We note that the results are shown for $\Omega_{m,0}$ and $H_0$ which are the parameters of the $f(R)$ model instead of $\Omega^\Lambda_{m,0}$ and $H_0^\Lambda$ (parameters of the $\Lambda$CDM model). These quantities are related by Eq.~\eqref{LCDMtoAlt} allowing us to obtain $\Omega_{m,0}$ and $H_0$ from $\Omega^\Lambda_{m,0}$ and $H_0^\Lambda$ by a postprocessing. 
For this, we also need the value of $H_0$ which is obtained evaluating for each model $H(z=0)$.

The posterior distributions and confidence contours of the cosmological parameters $\Omega_{m,0}$ and $H_0$ are very similar for both models analyzed in this paper. Therefore, we offer next the corresponding discussion that applies to both models. We note that the obtained value of $\Omega_{m,0}$ is consistent with the ones inferred considering other cosmological datasets \cite{Planckcosmo2018,DES} and assuming the $\Lambda$CDM model. Furthermore, the obtained value of $H_0$ is consistent with the value estimated by Ref.~\cite{Riess2022} using data from SNIa and Cepheids from the SH0ES Collaboration. This result is not surprising since we are using the same datasets: SNIa data together with the prior on $M$ that is derived from the SH0ES data (see discussion above). The only difference is that the analysis performed by Ref.~\cite{Riess2022} is completely model independent, while we are assuming an alternative cosmological model based on modified gravity theories. On the other hand, the obtained confidence interval of $H_0$  in Ref.~\cite{f_R_with_custom_z_0} [for both Hu-Sawicki and  Starobinsky $f(R)$ models],  when  the SH0ES data are not included in the analysis, is consistent with the one inferred from CMB data~\cite{Planckcosmo2018} and in tension with the one obtained by Ref.~\cite{Riess2022}. Therefore, we verify here, as discussed elsewhere, that the high values of $H_0$ are only obtained when the SH0ES data are considered.

Now, we focus on the posterior distribution of the distortion parameter $b$ for the Hu-Sawicki model. We note that the obtained confidence intervals of the $b$ parameter are shifted to larger values with respect to the one obtained in our previous work for the same model and the Pantheon compilation. This is the same behavior as the one obtained with the Starobinsky model in this paper and also in Ref.~\cite{f_R_with_custom_z_0}.

Next, we discuss the posterior distribution of the $b$ parameter for the Starobinsky model. We observe significant variations in the confidence intervals for parameter $b$ based on the choice of datasets used for the statistical analysis. Specifically, the confidence interval for $b$ obtained here with the Pantheon+ compilation data differs notably from the values reported in Ref.~\cite{Farrugia2021}, which used the Pantheon Collaboration data together with other datasets. On the other hand, the obtained confidence interval of $b$ in Ref.~\cite{f_R_with_custom_z_0}, where the Pantheon+ compilation data are considered, has also larger values than the ones obtained in Ref.~\cite{Farrugia2021}, as well as larger values than the ones obtained here. The reason for this last discrepancy can be found in the choice of the prior on $M$. 
We recall that a large value of $b$ in the Starobinsky model does not imply a huge departure from $\Lambda$CDM as is the case for the $f(R)$ Hu-Sawicki model \cite{Farrugia2021}.

Our results and the ones of Ref.~\cite{f_R_with_custom_z_0} (for both the Hu-Sawicki and the Starobinsky models), point out that the data from the Pantheon+ compilation are more consistent with departures from the $\Lambda$CDM model than previous datasets from the same collaboration.
We also noted an absence of degeneration in the $H_0\text{-}b$ plane (in both models) and this behavior is also observed in Ref.~\cite{f_R_with_custom_z_0} where a different choice for the prior on $M$ was made. Therefore, we conclude that the negligible correlation between $H_0$ and $b$ is not affected by the choice of the prior on $M$. 
This also discards the possibility that the larger value of $b$ allowed by the Pantheon+ dataset is rooted in any degeneracy between $b$ and $H_0$. Furthermore, we note a degeneration in the $\Omega_m\text{-}b$ plane in Fig.~\ref{contours_Starobinsky} as was also observed in Ref.~\cite{f_R_with_custom_z_0}.

\begin{table*}[]
    \centering
    \begin{ruledtabular}
\begin{tabular}{cccccc}
           & $b$                    & $\Omega_{m,0}$         & $H_{0}\left(\dfrac{\mathrm{km/s}}{\mathrm{Mpc}}\right)$ & $M$ & ${\chi}_{\nu}^2$ \\ \hline
$68\%$ C.I. & {[}0.255, 1.465] & {[}0.216, 0.291] & {[}70.582, 72.662]                                & {[}-19.327, -19.267] &               \\
$95\%$ C.I. & {[}0, 2.049] & {[}0.181, 0.321] & {[}69.572, 73.682]                                 & {[}-19.357, -19.239] &               \\
Mean   & 1.018                  & 0.251                  & 71.613                                                  & -19.298 &              \\ 
Best fit   & 0.296                  & 0.306                  & 67.883                                                  & -19.416 & 0.876             \\ 
\end{tabular}
\end{ruledtabular}

    \caption{Constraints on the parameters of the $f(R)$ Hu-Sawicki cosmological model with $n = 1$. The table shows the
$68 \%$ and $95 \%$ confidence intervals (C.I.) for each free parameter, along with their mean values. The reduced $\chi_\nu^2$ is also shown for the best fit. }
    \label{table:mcmc_results_HS}
\end{table*}

\begin{table*}[]
    \centering
    \begin{ruledtabular}
\begin{tabular}{cccccc}
           & $b$                    & $\Omega_{m,0}$         & $H_{0}\left(\dfrac{\mathrm{km/s}}{\mathrm{Mpc}}\right)$ & $M$ & ${\chi}_{\nu}^2$ \\ \hline
$68\%$ C.I. & {[}0.872, 2.706] & {[}0.252, 0.304] & {[}70.551, 72.636]                                & {[}-19.329, -19.269] &               \\
$95\%$ C.I. & {[}0.135, 3.157] & {[}0.226, 0.328] & {[}69.534, 73.63]                                 & {[}-19.36, -19.24] &               \\
Mean   & 1.714                  & 0.277                  & 71.588                                                  & -19.299 &              \\ 
Best fit   & 0.753                  & 0.312                  & 67.842                                                  & -19.417 & 0.876             \\ 
\end{tabular}
\end{ruledtabular}

    \caption{Constraints on the parameters of the $f(R)$ Starobinsky cosmological model with $n = 1$. The table shows the
$68 \%$ and $95 \%$ confidence intervals (C.I.) for each free parameter, along with their mean values. The reduced $\chi_\nu^2$ is also shown for the best fit. }
    \label{table:mcmc_results_ST}
\end{table*}

\subsection{Improved computational efficiency}
\label{sec:improved_comp}

One of the main motivations behind this work is to improve the speedup achieved on inference processes with the NN bundle method in our previous work.

In this section, we show our results regarding the time that it took to complete different MCMC processes, which use the same datasets, depending on the method used to compute the solutions of the given model being tested. These methods include a numerical method (Runge-Kutta), the NN method of our previous work, and the NN method presented here. For the NN method corresponding to our previous paper, we run all MCMC processes on CPU as well as on GPU. On the other hand, we run only on CPU for the method introduced here (see discussion below for more details). For the numerical method,\footnote{We recall here that we used the following Runge-Kutta methods from the SciPy library: RKF45 for the Hu-Sawicki model and Radau IIA family of order 5 for the Starobinsky model. We verified that using Radau IIA for the Hu-Sawicki model results in an increase of the computational times, while using it for the Starobinsky model implies an important reduction.} we set the error tolerances to $r_{\rm tol}=10^{-3}$ and $a_{\rm tol}=10^{-6}$, which are both the default values in SciPy's \cite{SciPy} method for solving initial value problems, and 
also match the order of the error in Figs.~\ref{fig:heatmap_Hu-Sawicki} and \ref{fig:heatmap_Starobinsky} (using the same ground truths as comparison). 
We perform these analyses for the Hu-Sawicki and Starobinsky $f(R)$ models.

\cref{table:time_table} shows our results for the MCMC algorithm computing times. In each column, we observe that the fastest method is the one presented in this work. Indeed, the new NN method moves the computational cost of both solving Eqs.~\eqref{diff_f_R_2} and computing Eq.~\eqref{dl_theo} to the training stage of the NNs. In this way, the computations needed at each step of the MCMC algorithm are just the evaluations of the necessary NNs on only the 1622 data points (cosmic chronometers and SNIa). 
Also, the fact that these points are not so many means that there is no incentive to leverage the parallelization capabilities of a GPU, due to the CPU being faster. We have corroborated that this is indeed what happens in our case. Nevertheless, this could be subject to change if the amount of data is large enough. This behavior is not the case for the old NN method.  In that case,  the parallelization that the GPU provides is far more important. The reason for this is the necessity of evaluating the NNs in a large number of points to perform the integration in Eq.~\eqref{dl_theo} (see Sec.~\ref{sec:method} for more details).

One more trend that can be observed from \cref{table:time_table} is that the advantage of using either NN method is larger for the Starobinsky model than the Hu-Sawicki model. This is because the Starobinsky model is more complex than Hu-Sawicki, which makes it more difficult to solve for the numerical method, and thus the advantage that either NN method provides over the numerical is enhanced. This further shows that, regardless of the improvements made here, a major factor on whether the numerical method or either NN method is faster, comes down to the difficulty of the problem that is being solved. 

On the other hand, let us mention some differences in the settings of the MCMC process of this work with respect to the ones of our previous work that affect the calculation of the respective computational times: (i) in this work we did not consider the times of postprocessing of the chains as was done in our previous work, (ii) the tolerance of the Runge-Kutta method is different, and (iii) the amount of iterations of the MCMC process also changed. A detailed discussion of all these changes can be found in \cref{apx:comp_times_details}.

There is an important clarification to be made regarding the time that it takes to compute the calculations of the MCMC algorithm using the numerical method for the Starobinsky model in \cref{table:time_table}. For that blue{run of the} MCMC algorithm we found that the numerical method struggled to compute solutions in certain regions of the parameter space [roughly $\left(b,\Omega_{m,0}^\Lambda \right)\in\left[3,4\right]\times\left[0.1, 0.17\right]$]. We estimate that this would make the time for the MCMC algorithm to complete on the order of a month. To mitigate this issue we adjusted the prior on $\Omega^\Lambda_{m,0}$ to be uniform such that $\Omega^\Lambda_{m,0}\in\left[0.17, 0.4\right]$, differing from the more natural prior of $\Omega^\Lambda_{m,0}\in\left[0.1, 0.4\right]$ we used for the NN method. We saw that this change in the prior did not seem to affect the results given the fact that the posterior distributions did not have a significant number of samples close to the boundary of $\Omega^\Lambda_{m,0}=0.17$. Nevertheless, because this choice of prior is not a natural one, we argue that the time we obtain is actually a lower bound on one corresponding to a more realistic scenario. We stress that we did not encounter this problem in either of the NN methods.

A relevant discussion to be had is the one corresponding to the time needed for the training of the NNs. The training that corresponds to Eqs.~\eqref{diff_f_R_2} was two days, while for Eq.~\eqref{eq:integral_to_diffeq} was 5 h 31 min, with both being done on a single NVidia A100 GPU. Even though these numbers seem to negate the advantage suggested by the ones shown in \cref{table:time_table}, we echo our arguments from our previous work in Appendix F 4 of Ref.~\cite{our_previous_paper}. There we stressed that, because the training of the NNs is done only once, the time taken for training becomes negligible over repeated use of the solutions.

While it is true that, generally, the training of the NNs is done once, it is important to remark on cases where additional training would be needed:
(i) a different theoretical model is considered to be tested against the data, (ii) new data that lay outside of the training range of the independent variable are considered, and (iii) the inclusion of new data results in confidence intervals whose range is outside the training range of the NNs.
This last example was the case when we tried to use  the NNs we trained for the  Hu-Sawicki model in our previous work \cite{our_previous_paper} for the statistical analysis performed in this work. Therefore, we trained new NNs with a wider training range in $\Omega^\Lambda_{m,0}$. These newly trained NNs cover a wide enough range so that we expect that there will be no need to retrain this networks in the future.

\begin{table}
\centering
\label{table:time_table}
\begin{ruledtabular}
\begin{tabular}{lll}
Method (hardware) & \begin{tabular}[c]{@{}c@{}}Hu-Sawicki\\model\end{tabular} & \begin{tabular}[c]{@{}c@{}}Starobinsky\\model\end{tabular} \\ \hline
Numerical method (CPU)   & 11 h 51 min       & > 2 days 17 h         \\
Old NN method (CPU)    & 17 h 21 min        & 2 h 25 min         \\
Old NN method (GPU)    & 11 h 51 min       & 1 h 33 min         \\
New NN method (CPU)    & 2 h 6 min        & 18 min         \\
\end{tabular}
\end{ruledtabular}

\caption{The computation times of MCMC processes applied to two different $f(R)$  models (Hu-Sawicki and Starobinsky) using the same datasets are compared across various techniques for computing solutions. These techniques include a numerical method (Runge-Kutta), the NN method described in our previous work~\cite{our_previous_paper}, and the NN method introduced in this study. All CPU computations were performed on an Intel i5-8400 and, for GPU computations, a single NVidia A100 was utilized.}
\end{table}

\section{Conclusions}
\label{conclusions}

In this work, we introduce the application of the NN bundle method to solve integral equations in scenarios where the integrand is computed with solutions of a differential system. 
We apply this approach to compute the luminosity distance, defined by an integral equation 
whose integrand is a function of the Hubble parameter which, in turn, is calculated by solving a differential system. By employing this technique, we develop a completely analytical likelihood function for use in an MCMC process aimed at constraining a cosmological model's parameters. This results in a notable reduction in the MCMC algorithm completion time compared to utilizing a numerical method or the NN method employed in our previous work \cite{our_previous_paper}.

During our study, we found that, for the Starobinsky $f(R)$ model, certain numerical instabilities reported in Ref.~\cite{f_R_with_custom_z_0} and replicated here are not present when using the NN method. In the numerical approach, these instabilities are typically addressed by modifying the initial conditions in a manner that increases the computational cost of solving the differential equations.

One of the main achievements of this work is the set of trained neural networks of alternative cosmological models that offer an improvement in speed of the MCMC inference process. Once trained, these networks can be used indefinitely with current and future datasets. Furthermore, the idea of computing the luminosity distance with the new NN method can be applied to other theoretical models to reduce the computational times of the inference process. Greater improvements in computational times will be obtained from using the new NN method for models that are computationally intensive to integrate.

On the other hand, this work provides an original analysis of the Hu-Sawicki and Starobinsky $f(R)$ models with recent data from cosmic chronometers and SNIa from the Pantheon+ compilation. The originality of our analysis lies in the appropriate treatment of the absolute magnitude of SNIa, as has been discussed in Sec.~\ref{subsubsec: methodology_stat}. This choice affects the estimation of the distortion parameter $b$ which measures the departure of the model with respect to $\Lambda$CDM (see discussion in Sec.~\ref{Results}).

In summary, we present here a novel way to incorporate the NN bundle method into an inference pipeline, making the likelihood analytical for any model whose observables are described by solutions to differential and integral equations. Consequently, the time required to complete the statistical analysis is significantly reduced.

Finally, further improvements  that can be brought into the method to make it an even more appealing option, are left for future work. Among them, we can mention:

\begin{enumerate}[label=(\roman*)]
    \item to improve the method so that it can provide the error of its own solution (this has been explored in Ref.~\cite{errorawarepinns}, but the application to the cosmological scenario is missing);
    \item optimization of the NN architecture to further improve the speed of the training stage, as well as the inference process; and
    \item perform the inference process with the Hamiltonian Monte Carlo algorithm \cite{HMC} instead of the MCMC algorithm (this can be achieved due to the fact that the likelihood now is analytical and, even more important, differentiable). 
\end{enumerate}

\begin{acknowledgments}
This collaboration was facilitated by AstroAI at the Center for Astrophysics | Harvard \& Smithsonian. The computations performed in this paper, which used GPUs, were run on the FASRC Cannon cluster supported by the FAS Division of Science Research Computing Group at Harvard University. S.L. is supported by Grant No. PIP 11220200100729CO CONICET and Grant No. 20020170100129BA UBACYT. C.G.S. is supported by Grant No. PIP 11220200102876CO CONICET, and Grant No. G175 from UNLP. C. G. was funded by NASA Contract No. NAS8-03060 to the CXC and by AstroAI. 
\end{acknowledgments}

\appendix

\section{Type Ia supernovae nuisance parameters}
\label{nuissance}

The distance modulus can be expressed
 by a modified version of the formula proposed in Ref.~\cite{Tripp},
\begin{equation}
    \label{mu_pantheon}
    \mu = m_b - M +\alpha x_1 -\beta c + \Delta_M + \Delta_B.
\end{equation}
Here $\alpha$ and $\beta$ are global nuisance parameters that relate stretch ($x_1$) and color ($c$), respectively, to luminosity. Additionally,  $M$  is the absolute magnitude of a SNIa.  $\Delta_B$ is a correction term that accounts for selection
biases (the procedure is described in detail in Appendix A of Ref.~\cite{PantheonPlus_data}). Finally,  $\Delta_M$ is the
luminosity correction for residual correlations between
the standardized brightness of a SNIa and the host galaxy, which depends on the nuisance parameter $\gamma$. This way, the corrected overall flux normalization is defined as $\tilde{m}_b=m_b+\alpha x_1 - \beta c +\Delta_{M} + \Delta_{B}$. The so-called nuisance parameters ($\alpha$, $\beta,\gamma$) are usually estimated from the comparison of the observational data with the standard cosmological model ($\Lambda$CDM). It has been shown that considering alternative cosmological models does not significantly change the estimation of these parameters \cite{Negrelli_2020,Matias_paper}. 
Therefore, for the purposes of testing cosmological models, the Pantheon+ compilation also provides a dataset where all nuisance parameters are fixed assuming the $\Lambda$CDM model and ,in consequence,  Eq.~\eqref{mu_pantheon} is reduced to Eq.~\eqref{mu_abrev}.

\section{Computing time details}
\label{apx:comp_times_details}
The main goal of our previous work \cite{our_previous_paper} was to test the viability and  advantages of the NN bundle method in the cosmological context. As part of that analysis, we reported the computing times of the inference process with a similar procedure to the one in Sec.~\ref{sec:improved_comp}. Nevertheless, given that the main goal of the work presented here is the advantage that the NNs provide in regard to the computing times, we have refined our methodology for the comparison between the numerical method and the NN method. For this, we have introduced two changes in the settings of the MCMC process.

First, we have increased the value of the error tolerances of the Runge-Kutta method in this paper to $r_{\rm tol}=10^{-3}$ and $a_{\rm tol}=10^{-6}$ from $r_{\rm tol}=10^{-6}$ and $a_{\rm tol}=10^{-9}$ of our previous work. This is more fair to the numerical method because it makes its precision comparable to the NN method while making the method faster. It is also important to recall that the tolerances used here are the default ones in SciPy's \cite{SciPy} method for solving initial value problems. 

The other change in methodology with respect to our previous work is that we no longer take into account the time needed to do the postprocessing of the chains required to obtain $H_0$ and $\Omega_{m,0}$ (for more details see Sec.~\ref{model}). While this computation is unavoidable, it is theoretically possible to optimize the MCMC algorithm so that it is no longer necessary to integrate the equations again to do the postprocessing. Because of this fact, it was unfair for the numerical method to include the post processing time because it meant to integrate again for each chain in order to obtain $H_0$, while for the NNs it just entailed evaluating them at $z=0$.

One last note on the difference between the times shown in \cref{table:time_table} and the ones shown in our previous work is the difference in the amount of iterations of the MCMC algorithm. Because we changed the data and priors for the Hu-Sawicki model (new cosmic chronometer data, updated Pantheon+ sample, the Gaussian prior for $M$, and the bounds on the prior of $\Omega_{m,0}^\Lambda$), the amount of iterations needed for the MCMC algorithm changed from our previous work. For example, the amount of iterations needed for the MCMC algorithm to converge for the Hu-Sawicki model in our previous work was 22000, while the one corresponding for this work was 206500. With regard to the Starobinsky model, the iterations needed were 28400. This is why the time it took for the MCMC algorithm to complete on the Hu-Sawicki model on the old NN method in \cref{table:time_table} does not match the one in our previous paper, even though the NN method is the same. Also,  this difference between the amount of iterations explains why the NN method took more time for the Hu-Sawicki model in \cref{table:time_table} than for the Starobinsky model. Otherwise, the process should take the same amount of time, because in both cases the computations are the same (i.e. evaluating three NNs on the same data points in each iteration of the MCMC algorithm).

\end{document}